\documentclass[pra,superscriptaddress]{revtex4}

\usepackage{amsmath}
\usepackage{graphicx}
\usepackage{dsfont}
\usepackage{epstopdf}
\usepackage{amsfonts}
\usepackage{color}
\usepackage{hyperref}
\usepackage{bm}
\usepackage{array}
\usepackage{tabularx}

\definecolor{darkblue}{rgb}{0,0,0.75}
\definecolor{darkred}{rgb}{0.5,0,0}
\definecolor{dg}{rgb}{0,0.3,0}

\hypersetup{
 pdfborder={0,0,0},
 colorlinks=true,
 linkcolor=darkblue,
 citecolor = darkblue
}

\begin{document}
\title{Mechanism of Laser/light beam interaction at cellular and tissue level and study of the influential factors for the application of low level laser therapy}

\author{Muhammad Zeeshan Khalid\footnote{Electronic address: zeeshan.khalid039@gmail.com}}
\affiliation{Department of Basic sciences, University of Engineering and Technology, 47050 Taxila, Pakistan }

\begin{abstract}
After the discovery of laser therapy it was realized it has useful application of wound healing and reduce pain, but due to the poor understanding of the mechanism and dose response this technique remained to be controversial for therapeutic applications. In order to understand the working and effectiveness different experiments were performed to determine the laser beam effect at the cellular and tissue level. This article discusses the mechanism of beam interaction at tissues and cellular level with different light sources and dosimetry principles for clinical application of low level laser therapy. Different application techniques and methods currently in use for clinical treatment has also been reviewed.
\end{abstract}

\maketitle

\section{Introduction and History}

Even after appearing first publication on Low Power Laser Therapy (LPLT) before 45 years ago, this topic remains the point of discussion for scientists to use this technique for clinical treatment. Eastern European Doctors in between 1960 to 1970 developed laser bio-stimulation, but it received tremendous criticism from scientist stating that; Low intensity laser radiation has direct effect on molecular level of organism. Even though some western supporters such as Italy, Spain, France as well as China \& Japan developed this method, but the method is still very controversial and need more solid understanding for implementing it into safe and reliable treatment method for clinical use. The much contentious points of laser biostimulation were extensively reviewed and analyzed in late 1980s \cite{karu1987photobiological, karu1989photobiology}. The some of the controversial points are the determination of ideal location of treatment of this method, dose rate, timing, wavelength and duration.

Since then some of the disputable points of LPLT has been addressed and some are being currently in the process of consideration of researchers and scientists. Currently LPLT is amalgamation of part of Physiotherapy and light therapy, it is also known as Low Level Laser Therapy (LLLT) or Photo-modulation. Light therapy was a very ancient method used by human centuries ago (also known as sun therapy or UV therapy historically). Use of laser and LED light is the new development method in the therapeutic field.
Laser therapy and Physiotherapeutic methods are very much alike due to the use of physical factors such as microwaves; Low frequency-pulsed Electromagnetic field, Focused ultrasound, Time-varying static and combined magnetic field, direct current etc. Biological Reponses of these physical features has been reviewed \cite{karu1988molecular}. Furthermore, recently different light sources (LED, laser and incoherent) for the treatment of wound healing was also reviewed on different biological targets \cite{suan2014light}.

LLLT has been shown useful in different studies. Found this method to be fruitful for the short term treatment of acute pain, neck pain, and chronic low back pain, tumor \cite{wheeland1993lasers, kov1974stimulation, neves2013effects}. Use of lasers for the treatment of hair loss was reviewed in 2008 to 2014, but there is not much evidence to support the effectiveness of this method. Experiments performed on mice showed application of laser light to the backs of shaven mice induced the shaved hair to grow more quickly than the unshaved mice \cite{mester1967effect}.
LLLT involves the application of laser light to tissues and cells. Due to the sensitivity of human tissue to high energy laser light and its controversial nature due to the poorly understood biochemical mechanisms, this treatment method is empirical. Furthermore, knowledge and selection of important parameters such as wavelength, power density, fluence timing of the applied light and pulse structure is important for the effective results. The literature suggested the optimal doses for the specific application to induce a response. This paper discusses the parameters and underlying physics and mechanism for low level laser therapy and also discusses the some applications techniques currently used for the treatment using this method.
\section{Elements to consider for Low Level Laser Therapy}

LASER is the acronym of “Light Amplification by Stimulated Emission of Radiation” and it possesses several unique properties which include monochromaticity, Coherence, directivity, low divergence and brightness \cite{mcguff1964surgical}. In order to understand the feasibility of lasers for LLLT, we need to know these unique properties of Laser beam and their effect on the performance of LLLT \cite{pascu2000laser}.

\subsection{Monochromaticity}

\begin{figure}[h!]\centering
    \includegraphics[scale=0.5]{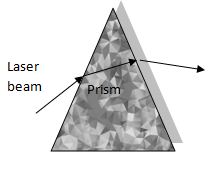}
    \caption{: LASER light through prism}\label{fig:prism1}
\end{figure}

Monochromaticity describes the spectral distribution of radiation, which is correlated to the intensity of light. Laser light is unique from ordinary light due to its one color, that’s why it is named as monochromatic (mono mean one, chromatic mean light). As shown in fig \ref{fig:prism1}, Laser light after passing through prism does not divide into different colors unlike the ordinary light.
So it is important to note in the case of laser, maximum intensity is obtained at a given wavelength at which most of the radiation emitted, the intensity falls to zero at other wavelengths other than maximum. This property is also named as “spectral purity of laser beam”.
Although; the property of monochromaticity is important for Laser beam, but in the case of LLLT this is not a very critical parameter due to the broad absorption bands of tissues. So there is not much need of spectrally narrow or stable wavelength output laser beam.
\begin{figure}[h!]\centering
    \includegraphics[scale=0.5]{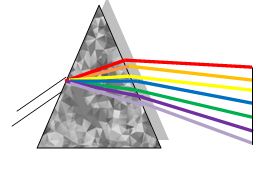}
    \caption{Ordinary Light through prism}\label{fig:prism2}
\end{figure}

\subsection{Coherence}

This property arises from the stimulated emission, since the emission of the laser beam is dependent upon the stimulus, which provides amplification, so during the stimulated emission; emitted beams have a phase relation to each other. Coherence is mainly characterized into Spatial and Temporal Coherence. Spatial coherence is the interference of Laser at two different points which may locate at a small or higher distance in laser cross section, while temporal coherence is a superposition of laser beam at certain points but in a different time line. So spatial coherence describes the field in the lateral direction so the divergence of light field is taken into consideration for spatial coherence as shown in equation 1.
\begin{equation}
\mathbf{l_c}\cong\frac{\mathbf{\lambda}}{\mathbf{\phi}}
\end{equation}

Where $\phi$ is the divergence of light beam at the point of irradiation.

Temporal coherence can be explained by spectral width   due to the protocol light oscillation during the coherent time $\tau_c$.

\begin{equation}
\tau_c\cong\frac{1}{\Delta v}
\end{equation}

Since the light beam travels with the speed of light $c=3\times10^8 m/s$, so the light is coherent on the length of light propagation is given as $L_c$.

\begin{equation}
\L_c\cong\frac{c}{\Delta v[Hz]}
\end{equation}
Equation $3$ shows the direct relation between coherent length and monochromaticity of light.
Concepts of temporal and spatial coherence are quite independent of each other. If say, for example, two beams are perfectly spatial coherence, but they will be limited temporally coherent or vice versa. So spatial and temporal coherence does not show  the higher order coherence properties.
It is still not clear, whether the coherence and monochromatic light have some bonus advantages in LLLT in comparison to the conventional or LED light with same center intensity and wavelength. Interaction of the laser beam with tissues or biomolecules produce the phenomenon of laser speckle, which is known for playing a part in photomodulation interaction with cellular organelles and cells \cite{rubinov2003physical}. That’s why it is  very challenging for scientists to design experiments for the direct comparison of coherence and non-coherence of laser light. Since laser light has a bandwidth of $1 nm$ or less so it is almost not possible to generate light from any other source to produce beam of wavelength around $10-20nm$.

\subsection{Directivity}

Directivity and divergence quite similar in a sense they both have same and equal effect on the laser beam. Directivity is defined in terms of divergence as the minimum divergence of a laser beam or uniformity of beam cross section along the small distances. Divergence of the laser beam is measured in the units of mill radians (mrad).Divergence describes quantitatively and Directivity qualitatively the concentration of laser beam along the long distances.  The shape of the active medium and the geometry of the cavity have a profound effect on the divergence of the laser beam. Plane Parallel Mirrors with infinite dimensions produces axial amplification, while there is a negligible transverse spatial effect in the active medium and resonator.
Spherical mirrors are used for experimental purposes since the plane resonator with typical dimensions in between $20mm$ to $50mm$ are very sensitive to mechanical misalignment, so in this way we can observe and determine the transversal modes with axial modes due to the increase in the divergence of the laser beam.
Mode structure is very important entity for the laser beam divergence and directivity characteristics. Divergence of laser beam differs from one mode to other. Gaussian mode $TEM_{00}$ is considered the ideal mode structure for laser beam, but real laser beam show fluctuations in the beam structure, so in order to produce perfect laser beam for LLLT application, it is important to study the laser modes in different geometries. General overview of the Laser modes is given in the next sub-section.

\subsubsection{Laser modes}

Laser beam are very much similar to plane waves, but their intensity distribution is concentrated to the axis of propagation. Coherent Laser beam field component follows the following scalar wave equation \cite{pascu2000laser}.

\begin{equation}
\Delta^2 u+k^2u=0
\end{equation}

k is the propagation constant $(k=2\pi/l)$

Light propagating in z direction
\begin{equation}
u=\Psi(x,y,z)\exp(-jkz)
\end{equation}
Where y is the wave function which slowly varies with the propagation of the laser beam with respect to the plane wave. It tells us about the curvature of phase front and expansion of the laser beam. Now by inserting equation $5$ into $4$ we get the general solution of laser beam propagation equation.
\begin{equation}
\frac{\partial^2 \Psi}{\partial x^2}+\frac{\partial^2 \Psi}{\partial y^2}-2jk\frac{\partial \Psi}{\partial z}=0
\end{equation}
Since the variation of the beam along z axis is assumed to be very slow, so the $2nd$ derivative is neglected along the z axis in equation $6$.
Solution of equation $6$ is given as:
\begin{equation}
\Psi=\exp{-j(P+\frac{k}{2q}r^2)}
\end{equation}
Where
\begin{equation}
r^2=x^2+y^2
\end{equation}

P(z) is the complex wave shift and describes the wave propagation of laser beam, while q(z) is a complex parameter describing the wave Gaussian variation in the beam intensity w.r.t. r and also tells us about the curvature of the phase front (Spherical near axis). By putting equation 7 in 6 and comparing equal terms of r we obtain q’=1 and P’= -j/q. Where q’ and P’ are the differentiation with respect to z. After integrating q’ and P’, we get equal 9, $q_2$ is the beam parameter in the output plane with beam parameter of input plane $q_1$ in the direction of z
\begin{equation}
q_2=q_1+z
\end{equation}
Equation 6 is the simple case of the beam whose intensity profile is given by a Gaussian distribution. The solution of this equation varies with different geometries and these solutions are orthogonal to each other and named as “Propagation modes”. Solutions of equation 6 in Cartesian and cylindrical coordinates are described below.

\subsubsection{Modes in Rectangular Coordinates}
For a rectangular system with (x, y, z) coordinates, we can take a trial solution for equation 6 in the form.
\begin{equation}
\Psi=g\frac{x}{w}.h\frac{y}{w}.\exp{(-j[P+\frac{k}{2q}(x^2+y^2)])}
\end{equation}
G and h are the function of x , z and  y, z respectively. w(z) is the scaling parameter of laser beam and it is applicable to all orders of modes. For real g and h value Gaussian beam has an intensity scale pattern of width $2w(z)$. After inserting equation 10 into 6, we get,
\begin{equation}
\frac{\partial^2 H_m}{\partial x^2}-2m\frac{\partial H_m}{\partial x}+2mH_m=0
\end{equation}
$H_m(x)$ is the Hermite polynomial of order m. Equation 6 is satisfied if we have,
\begin{equation}
g.h=H_m(\sqrt{2}\frac{x}{w})H_n(\sqrt{2}\frac{y}{w})
\end{equation}

\begin{figure}[t!]\centering
    \includegraphics[scale=1.0]{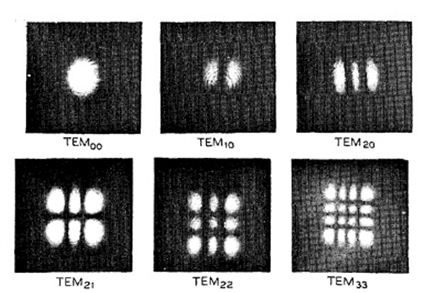}
    \caption{Transverse Mode patterns in Cartesian coordinates of Gas laser oscillator with m=0, 1, 2, 3 and n=0, 1, 2, 3 \cite{kogelnik1966laser}.}\label{fig:prism3}
\end{figure}
Some low orders Hermite Polynomials are:
\begin{equation}
  \begin{aligned}
  H_0(x)&=1\\
  H_1(x)&=x\\
  H_2(x)&=4x^2-2\\
  H_3(x)&=8x^3-12x\\
  \end{aligned}
  \end{equation}
  Product of Hermite Polynomial and Gaussian functions is used to describe the intensity pattern of higher order cross section beam. Figure \ref{fig:prism3} is the photographs of mode pattern in a gas laser oscillator.These modes can be generally described as $TEM_m,n,q$ the pair of (m,n) is the mode number, area of a mode increases with value of mode number.  (0,0) mode in the figure is characterized as Gaussian mode m and n shows the dark spaces between the spots in laser beam cross section horizontally and vertically. For a given pair of m and n, q shows the number of compatible longitudinal modes, which have different frequencies ($v_f = \frac{c}{2d}$, d=spacing between resonator mirrors), but the same spatial modes irrespective of the values of m and n.
  \subsubsection{Modes in Cylindrical coordinates}
  Trial solution of equation 6 for cylindrical coordinates $(r,\Phi,z)$ is given as:
  \begin{equation}
  \Psi=g(\frac{r}{w}.exp{(-j[P+\frac{k}{2q}(r^2+l\Phi)]))}
  \end{equation}
  We can find the value of g after some calculation
  \begin{equation}
  g=(\sqrt{2}\frac{r}{w})^l.L_p^l(2\frac{r^2}{w^2})
  \end{equation}
  $L_p^l$ = generalized Laguerre polynomial with p and l generalized radial and angular modes.

  $L_p^l(x)$  follows the differential equation.
  \begin{equation}
  x\frac{\partial^2 L_p^l}{\partial x^2}+(l+1-x)\frac{\partial L_p^l}{\partial x}+pL_p^l=0
  \end{equation}
  And some low order polynomials are given as
  \begin{equation}
  \begin{aligned}
  L_0^l(x)&=1\\
  L_1^l(x)&=l+1-x\\
  L_2^l(x)&=\frac{1}{2}(l+1)(l+2)-(l+2)x+\frac{1}{2}x^2\\
  \end{aligned}
  \end{equation}
  Beam parameters $w(z$) and $R(z)$ remain the same for Cartesian and cylindrical coordinates, phase shift depend on the mode numbers and is given below for both rectangular and cylindrical modes

  For rectangular modes:
  \begin{equation}
  \Phi (m,n;z)=(m+n+1)arctan(\frac{\lambda z}{\pi \omega_0^2})
  \end{equation}
  For Cylindrical modes:
  \begin{equation}
  \Phi (p,l;z)=(2p+l+1)arctan(\frac{\lambda z}{\pi \omega_0^2})
  \end{equation}
  \subsection{Polarization}
\begin{figure}[t!]\centering
    \includegraphics[scale=0.2]{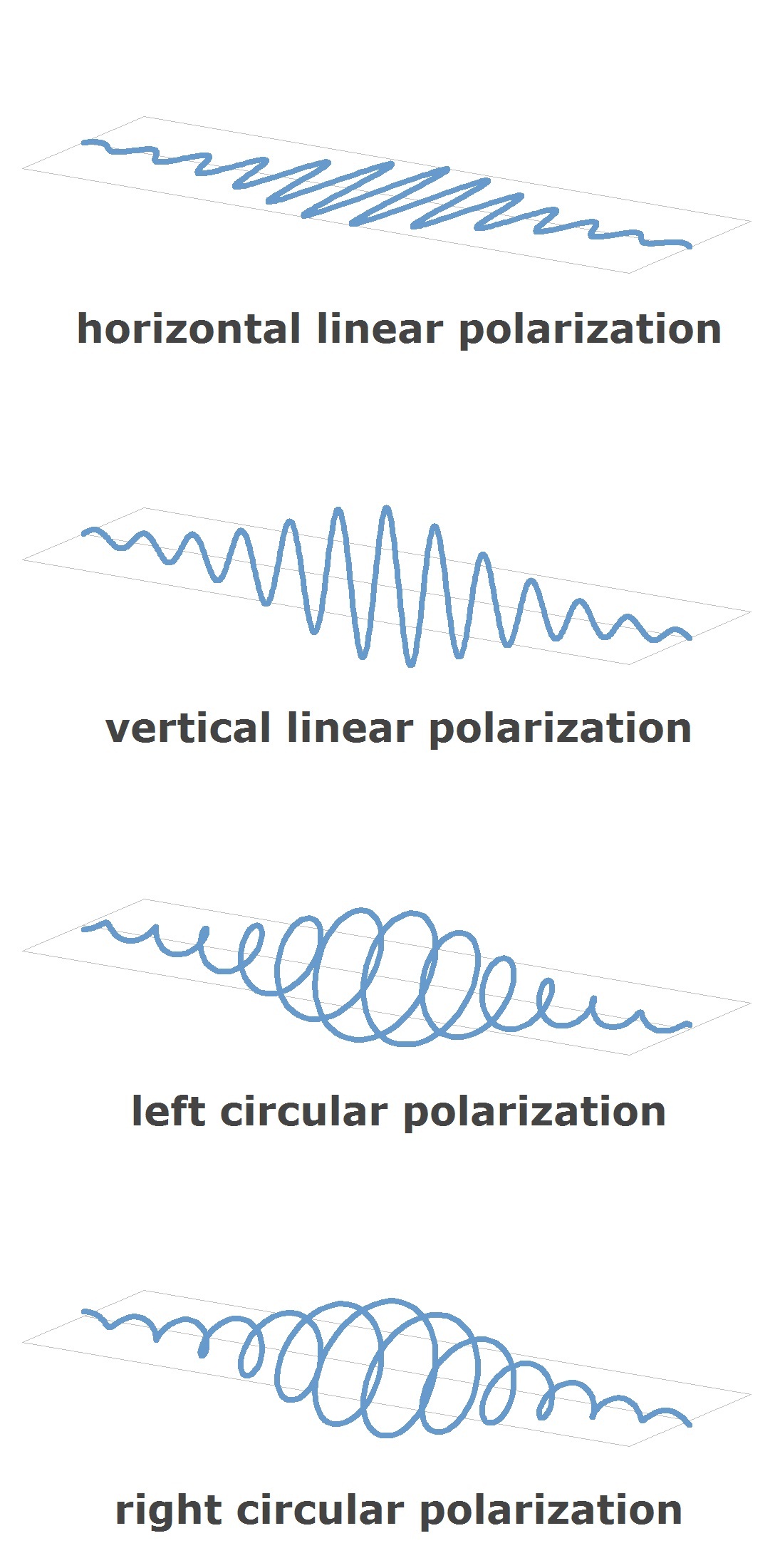}
    \caption{Polarization states of laser beam:  figure was reproduced from the Encyclopedia of Laser Physics and Technology and is used with permission by RP Photonics Consulting GmbH}\label{fig:prism4}
\end{figure}

  The laser beam is sometime is polarized which is the nonspecific property of Laser beam. Since this property of Laser coexist with other properties of Laser radiation, so it is important to know about the polarization state of Laser radiation. Laser radiation can be polarized in a specific way (Horizontal, Linear polarization, etc.) by the use of a polarizer. Figure \ref{fig:prism4} illustrates the different polarization states of laser radiation for a beam of few pulse cycles propagating from left to right.

In the previous section we introduced Transverse Electro Magnetic modes TEM m, n, q. Each mode (m,n,q) of emitted radiation have independent orthogonal polarization, so each oscillating mode corresponds to polarization of laser radiation.

Due to the circular symmetry of the two polarization mode of spherical mirror resonator, they have the same spatial coherence for the same pair (m, n). If the losses and gains of optical cavity and laser active medium are same for both polarizations then the laser beam will be un-polarized and will contain of two independent modes of equal intensity. Similarly we can get linear polarization if active medium and mirror of optical cavity accept one of the polarization states.

So it can be inferred for some anisotropic and semiconductor lasers, laser gains and resonator losses are polarization dependent. A stable linear polarization can be achieved by small gain or loss difference of two polarization direction, unless laser resonator has no coupled polarization modes.

\subsection{Astigmatism}
\begin{figure}[t!]\centering
\includegraphics[scale=.9]{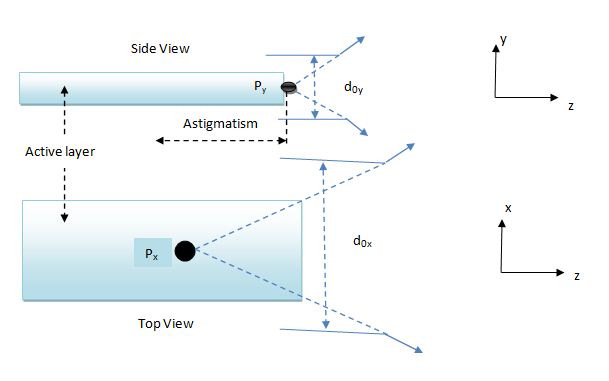}
\caption{Origin of Astigmatism of laser beam Top and side view \cite{pascu2000laser}.}\label{fig:prism5}
\end{figure}

Astigmatism is the phenomenon occurring when vertical and horizontal cross sections of beam focus are at different beam position (Fig.\ref{fig:prism5}) \cite{pascu2000laser}.
In fig.5 Px is the origin point of horizontal beam which is located behind Py (origin of vertical beam point). These two points have indirect relation to each other; the larger the dx will result in narrower of dy beam resulting in higher the distance between $P_x$ and $P_y$.

For LLLT this factor is very troubling for practical irradiation of tissues. This factor leads to the partially collimated beam which may not be very useful for clinical purposes. Full divergence of the angle of the emitted beam from a small aperture is given as
\begin{equation}
\theta=\frac{4\lambda}{\pi d}\
\end{equation}
Where $\lambda$ is the wavelength of emitted beam and d is the facet size of emitted beam.

Due to their inverse correlation, it is difficult to achieve collimation of the beam simultaneously in both the vertical and horizontal direction since $P_x$ and $P_y$ cannot be placed at the focal point of the collimating lens at the same time. The astigmatism has an important role in LLLT, since the beam collimating limit of laser diode, shape focus and dimension of the optical system can be controlled. Furthermore astigmatism can be very useful tool for other types of laser such as solid state ones which are optically pumped by diode lasers.
\subsection{Output power of laser}
\begin{figure}[h!]\centering
\includegraphics[scale=0.8]{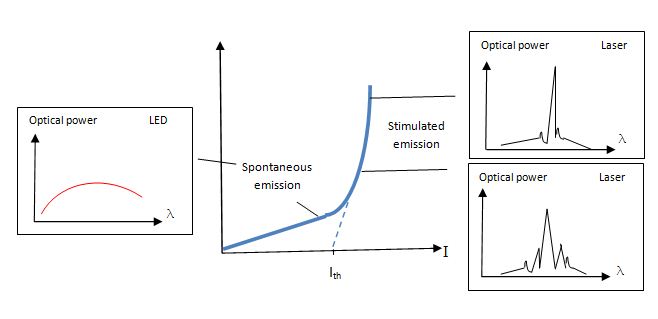}
\caption{Output optical power vs. diode current curve \cite{safa2009optoelectronics}.}\label{fig:prism6}
\end{figure}

Output power of the laser beam is very crucial factor of laser beam characteristics for the use in LLLT. Forward current intensity can be varied up to $100mA$ cw for LLLT application for diode laser working under injection current condition. From fig \ref{fig:prism6} it can be seen there is some threshold limit defined as Ith above which laser light is emitted; below this limit laser diode emit incoherent light by spontaneous emission. So the output power depends upon the injection current value, and the slope in graph describes the slope efficiency of laser light.

The maximum forward tolerable current for the production of optical power output Po under continuous operation is defined as the sum of threshold current value and the specified current of laser diode. Above this limit laser output is drastically affected due to the probable malfunctioning of laser machine. Averaged output power of pulsed laser regime can be calculated u sing equation 21 \cite{pascu2000laser}.
\begin{equation}
P_a(W)=Pp(W)\times \Delta \tau (s) \times f(s^{-1})
\end{equation}
Pp is expressed in unit of watts and defined as the power per pulse; $\Delta t(s)$ is pulse time width in second for a pulse f.

\section{Laser beam interaction with Tissues, Cells and biomolecules}[b!]
Laser beam interaction at molecular level shows an incoherent behavior of the laser beam. This problem was first described several years ago, this bizarre laser beam interaction with Biomolecules raised a question whether coherent light is needed for “Bio Stimulation” or it is simply a biological phenomenon. The study suggested incoherent light absorption (i.e. Photobiological) nature of low intensity laser beam under physiological condition due to the high rate of de-coherence of excitation than the rate of photo-excitation. Under normal condition it takes 10-12s for de coherence of photo-excitation, and this time depends upon the intensity of the laser beam (1mW/cm2 intensity takes about 1s).  The intensity for condensed matter compounds at which the interaction between matter and coherent light occurs was estimated well above GW/cm2 at 300K1. While for clinical purposes light intensity is not higher than tens of hundreds of mW/cm2. Experimental data in the literature suggested at the cellular level coherent and non-coherent light (LEDs, etc.) with the same wavelength, intensity and irradiation provides the same biological effect \cite{karu1982biostimulation, karu1983stimulation, bertoloni1993biochemical}.

\begin{figure}[h]\centering
\includegraphics[scale=0.9]{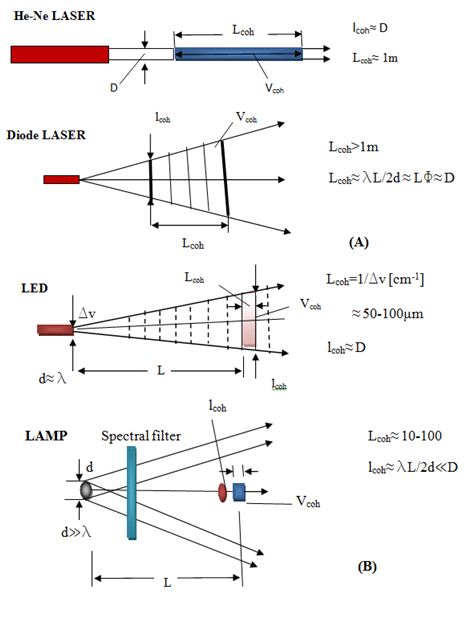}
\caption{Comparison between coherent and non-coherent radiation sources to study coherence volume and coherence length effect (A) Monochromatic light (B) conventional light Source \cite{karu2003low}.}\label{fig:prism7}
\end{figure}

Since the light coherence was not manifested at the cellular level so it seems possible to achieve coherence at macro level such as tissues of organs. Consider fig \ref{fig:prism7} which provides four different light sources, (A) two coherent light sources (Typical He-Ne and Diode lasers) (B) Two non-coherent light source (LED and filtered lamp). From fig \ref{fig:prism7} it can be seen Length of coherent is comparably very small for non coherent light source 7b than for Laser light, so large volume of tissues is irradiated when the light source is a monochromatic light \cite{karu2003low}.

Length of longitudinal coherence is more important than the spatial coherence, in spatial coherence the light is scattered when interact with bio tissues when the penetrating depth is much greater than the scattered coherent beam ($L \gg l_{coh}$), this is due to the fact every region is irradiated by radiation with an angle ($\Phi=1 rad$). So the light coherence varies with the wavelength of light lcoh=l. Longitudinal coherence length has an important role in the determination of the volume of the irradiated tissue, $V_{coh}$, when the bulk tissue is irradiated. This volume constitutes interference of scattered light and non-homogeneities of intensity. Coherence length is very small for non-coherent light source which is round about tens to hundreds of few microns which is quite low as compare to the coherent light source. So the length of coherent light source is not the only therapeutic effect which exists for laser source, but penetration depth of tissue caused by scattering and absorption has also additional effect for coherent light source (Table 1).
\begin{table}[h]
\caption{Light sources comparison of coherence for clinical and Experimental work \cite{karu2003low}.}
 \scalebox{0.8}{
\begin{tabular}{|c|c|c|c|c|>{\centering}m{5cm}|}
\hline
\bf{Light source}      & \bf{Temporal coherence} & \bf{Length of longitudinal coherence $\L_{coh}$} & \bf{Spatial Coherence} & \bf{Volume of Lateral Coherence, $l_{coh}$} \\
\hline
Laser             & Very High & Very Long & Very high & Large \\
LED               & Low       & Short($\gg \lambda$) & High & Small \\
Lamp with filter  & Low       & Short($\gg \lambda$) & Very low & Very small \\
Lamp              & Very low  & very short & very low & Extremely small \\
\hline
\end{tabular}}
\end{table}

\begin{figure}[t!]\centering
\includegraphics[scale=0.9]{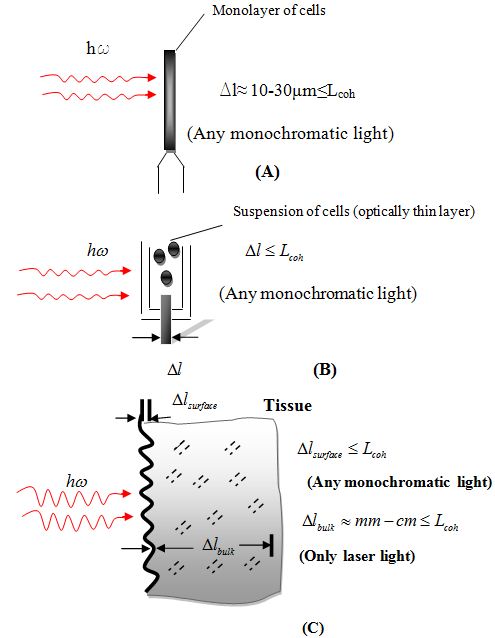}
\caption{Beam coherency manifestation at different depths in different irradiated systems (A) Monolayer of cells (B) suspension of cells (C) Tissues \cite{karu2003low}}\label{fig:prism8}
\end{figure}

The importance coherence length difference begins to fall when thin layers are irradiated with conventional light source (LED light or filtered lamps). Figure \ref{fig:prism8} illustrates the example of a monolayer of cells (A) and optically thin layers of cells (B). Experiment illustrated at figure \ref{fig:prism6} showed the same biological Reponses for the coherent and non-coherent light sources at the same parameters, but the situation become different for the case in figure \ref{fig:prism8}C when the bulk tissue is irradiated. Non-coherent light has very short coherent length which can play some role on the surface length ($Dl_{surface}$) of irradiated tissue. While the coherent light source maintained its coherent length along the entire length of bulk tissue thus indicating the great penetration power of laser beams \cite{karu2003low}.

Penetrating radiation shows random interference pattern over the entire bulk length ($Dl_{bulk}$), thus causing different speckle intensity pattern along bulk tissue, which cause maximum intensity pattern at constructive interference. The speckle or coherence effect appears only at the depth of $L_{coh}$. These speckles appear non-homogeneously in the bulk material, so the energy deposition is different thus resulted in statically non-homogenous photochemical processes such as increase in temperature, pressure, deformation of cellular membranes etc.
In order to increase the appearance of region with zero speckle intensity, polarizer is employed. Polarized coherent light reported to increase the manifestation of the effect of light coherence in irradiated tissue.

Directivity of laser beam play major role in beam penetration power. Its role become more prominent for some tissues which have fiber like structure, in this case waveguide propagation effect appear providing the more penetration power to laser beam.
We can conclude, in order to produce therapeutic effect for clinical purposes it is important to create the laser beam which will make it able to get absorbed by photoreceptor. Coherent and non coherent at the same parameters produce the same biological effect at micro level such as cell monolayer, layers of cell suspension, but coherent beam showed the penetration ability at macroscopic level.
\section{Mechanism of Low level laser therapy}
\subsection{First Law of Photobiology}
First Law of photobiology states that in order to have any effect by Low power light beam, photons must be absorbed by electronic bands of molecular photoreceptors or chromophores \cite{sutherland2002biological}. Chromosphores are defined as part of molecule which imparts its parent color to the compound. Examples of such chromophores are chlorophyll (imparts green color to plant), hemoglobin (blood cell), Cytochrome c oxydase (Cox), flavins, myoglobin, flavroprotein and porphyrins \cite{sutherland2002biological, kato1981cytochrome, passarella1984increase, gordon1960red}.
\subsection{Optical Properties of Tissues}
\begin{figure}[t!]\centering
\includegraphics[scale=0.4]{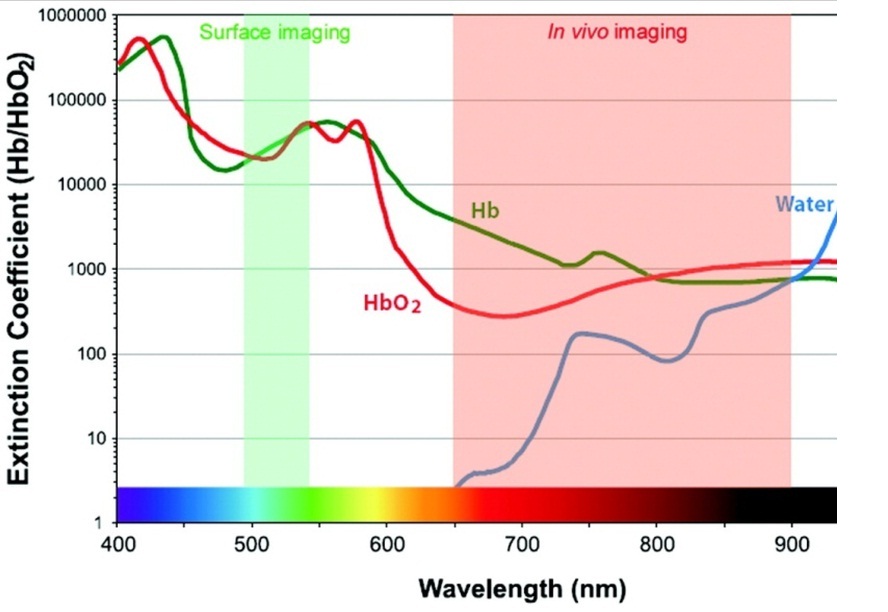}
\caption{Different chromophores showing range of wavelength of maximum penetration of light (Optical Window)-Reproduced under creative common license \cite{joshi2010exogenous}.}\label{fig:prism9}
\end{figure}

Optical properties of tissues are also one of the most important key aspects of LLLT. Absorption and scattering of light from tissues is mainly dependent upon wavelength of light (higher in blue region of spectrum than red) \cite{hamblin2006mechanisms}. Chromophores (melanin and hemoglobin) have absorption power at wavelength of shorter than 600 nm and water at 1150 nm. Different studies in literature proved the effectiveness of laser beam at wavelengths of 415 \cite{kato1981cytochrome}, 602 \cite{vekshin1981flavin}, 632.8 \cite{passarella1984increase}, 650 and 725 nm \cite{gordon1960red} by increasing the ATP production, but wavelength beam at 474 and 554 nm \cite{kato1981cytochrome} did not show any influence on the process. Different cells and elements have different wavelengths for absorption of light as shown in fig \ref{fig:prism9}, that’s why there exists a “Optical Window” in tissues at which penetration of light is maximized. Due to the existence of optical window in between red and NIR wavelength range, use of LLLT in animals and patients exclusively involves at this wavelength range of light \cite{karu1990effects}.During the clinical application light partly absorbed and partly reflect by biological tissues can be explained by Snell’s law. Light reflection is produced by change in air and tissue refractive index as described in equation 22.
\begin{equation}
\frac{Sin \theta_1}{Sin \theta_2}=\frac{n_2}{n_1}
\end{equation}
  is angle surface to air and   is angle between ray surface normal to tissue, $n_1$ and $n_2$ represents the index of refraction at air and tissue respectively.

Energy absorption by tissues is characterized as the absorption coefficient $\mu_a (cm^{-1})$. This phenomenon is the main cause for the effects on the tissue. The inverse of absorption ($I_a$) defines the mean free path of light into the absorbing medium.

The other most important phenomenon is the scattering of light because it determines the light intensity distribution in the tissue. We can define scattering coefficient in the same way as defined absorption coefficient and is denoted as $\mu_s$ ($cm^{-1}$).Mean free path length until the next event is defined by the inverse of scattering coefficient ($\frac{1}{\mu_s (cm)}$). Sum of the absorption and scattering coefficient is defined as the attenuation.
\begin{equation}
\mu_t=\mu_a+\mu_s
\end{equation}
At cellular and Tissular level scattering is not uniform, predominating the forward scattering. That’s why different anisotropic values g.g from 0 to 1 are assigned (for isotropic g=0, forward scattering g=1). Value of g in tissue can have value from 0.8-0.99. By considering this value reduced scattering coefficient ($\mu_s(cm^{−1}$) is introduced and defined as;
\begin{equation}
\mu_s=\mu_s(1-g)
\end{equation}
\subsection{Transport theory}
Transport theory is used to describe the transport of light energy at Tissular level. According to this theory absorption and scattering reduce the radiance L(r,t) of light at position r along the unit vector s, but it increase for light which scattered from s′ to s direction Radiance describe the light emitted, reflected, transmitted or received from a surface area per unit solid angle in a specified direction Light interaction according to this theory can be described as:
\begin{equation}
s.\Delta L(r,z)=-(\mu_a+\mu_s)L(r,s)+\mu_s \int_{4\pi}p(s,s')L(r,s')d\omega'
\end{equation}
p (s, s′) is phase function and   is the solid angle in s′ direction [39].

It is often very difficult to solve transport equation that’s why phase function and radiance are approximated and approximate solution depend upon the type of  light irradiation (collimated or diffuse) and optical boundary condition \cite{cheong1990review}.
\subsection{Cellular and Tissular Effect of Radiation in LLLT}
The main power source at cellular level is Mitochondria also known as the “Cellular Power house”. It plays an important role in energy generation and metabolism by converting food molecules into Adenosine triphasphate (ATP) during the process of oxidative Phosphorylation. From observations it’s been suggested LLLT acts to increase the production of ATPs, modulation of reactive oxygen species and in the induction of transcription factor by acting on the Mitochondria \cite{chen2011low, passarella1983evidence, karu1995irradiation, bakeeva1993formation, manteifel1997ultrastructural, herbert1989effect, lovschall1998low, anders1995low}.These transcription factors further induce protein synthesis, cell proliferation \cite{yu2003helium}, migration \cite{hawkins2005low}, cytokines modulation, growth factor and inflammatory mediators and increase tissue oxygenation \cite{chen2011low, yu2003helium, song2003cdna}.

\begin{figure}[b!]\centering
\includegraphics[scale=0.7]{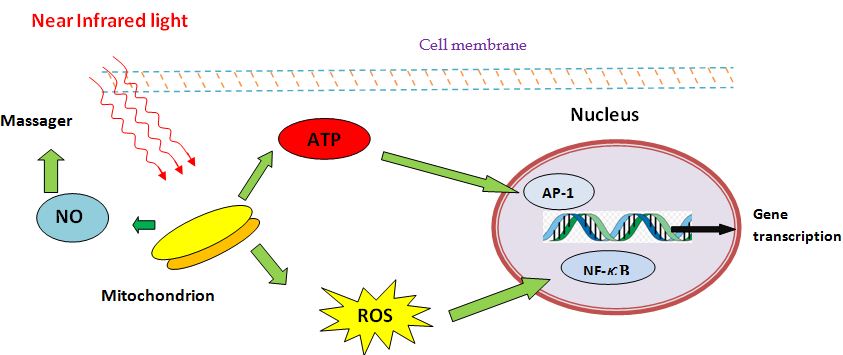}
\caption{Cellular and molecular mechanism of LLLT \cite{chung2012nuts}}\label{fig:prism10}
\end{figure}

Fig \ref{fig:prism10} explain the LLLT mechanism during which Near Infrared light or red light absorbed by chromophores or photoreceptor in the mitochondria which triggers oxygenation producing more ATPs, resulting in the production of ROS and generation of Nitric Oxide (NO) . These effects induce transcription changes by activating transcription factors. Some of the transcription factors which produce such responses are nuclear factor keppa-B (NF-κB), activating transcription factor (cAMP), response element binding protein (ATF/CREB), hypoxia-inducible factor (HIF)-1and HIF-like factor \cite{chen2011low}.

\begin{figure}[t!]\centering
\includegraphics[scale=0.7]{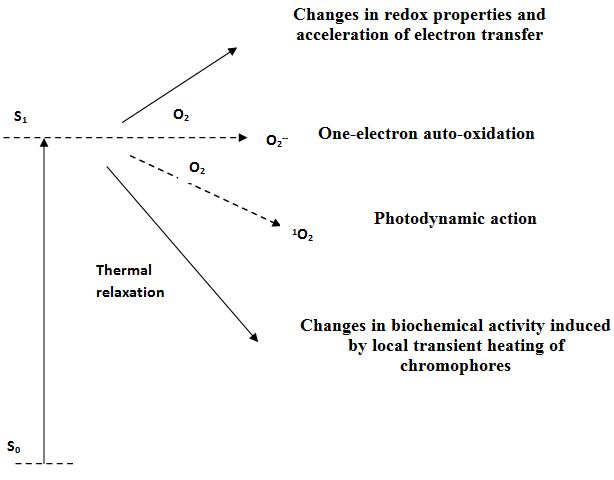}
\caption{Possible reactions due to photo excitation of electron \cite{lubart1992effects}.}\label{fig:prism11}
\end{figure}

\begin{figure}[b!]\centering
\includegraphics[scale=0.7]{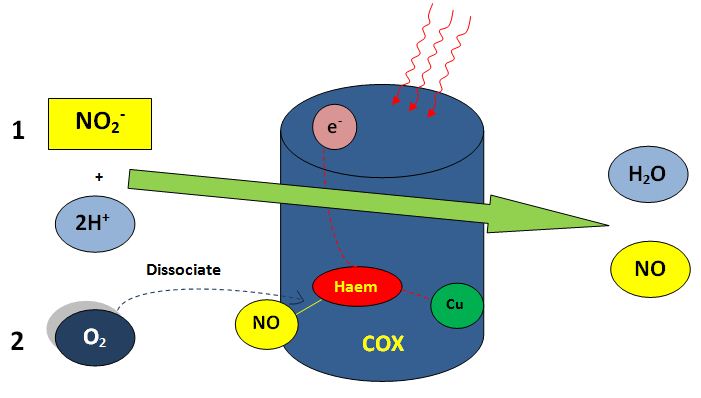}
\caption{Two possible mechanism of NO release from CCO (1) Reduction of NO by nitrate reductase enzyme (2) Photo-disassociation of NO from CCO \cite{chung2012nuts}}\label{fig:prism12}
\end{figure}

Visible or near IR light brought changes in cells due to physical or chemical changes in photo respiratory molecules or components of reparatory change. After the absorption of light electrons after absorbing photon energy excite to higher states, as a result of photo excitation there are following probabilities of occurring physical/chemical changes in cells (Fig. \ref{fig:prism11}).
\begin{itemize}
\item Acceleration of electron transfer or alteration of redox properties.
\item Local transient heating of chromophores induce changes in biochemical activity.
\item $O_2$ production and one-electron auto oxidation.
\item Photodynamic action and $O_2^1$ production.
\end{itemize}
The study found Cytochrome c oxidase (CCO) is the main chromophores which responds to the LLLT light \cite{karu1995cytochrome}. It is a cellular transmembrane which consists of two copper centers and two heme-iron centers \cite{capaldi1990structure}. These are the components of respiratory electron transport chain which passes high energy electrons through CCO and transmembrane complexes to the final electron acceptor thus producing more ATPs. In this way LLLT light influences transmembrane complexes and results in the production of more ATPs \cite{karu1995irradiation, pastore1994increase}.
There are two possible manners in which CCO and NO are released due to LLLT light (fig.\ref{fig:prism12}).

{Path 1}: A low partial pressure CCO acts as nitrate reductase enzyme at wavelength of $590 \pm 14 nm$ LED light under hypoxia condition \cite{ball2011low, lohr2009enhancement, zhang2009near, poyton2011therapeutic, ball2011low}.
$$NO_{-2}+2H^+ +e^-(CCO) \longrightarrow NO+H_2$$
Path 2: Disassociation of NO from CCO caused by LLLT \cite{karu2005cellular, lane2006cell}. NO decreases the ATP production by displacing oxygen from CCO and thus inhibiting the cellular respiration \cite{antunes2004mechanism}. This process can be prevented by LLLT which disassociate NO from CCO thus producing more ATP.
Role of LLLT does not end at increasing the ATP production, But oxygen which acts as the last electron acceptor thus producing water molecules and reactive oxygen species (ROS) by metabolism as a natural by-product. LLLT increase the cell metabolism thus increasing the ROS production which is chemically very reactive and influence the cell signaling, regulation of cell progression, nucleic and protein synthesis and enzyme activation. Thus LLLT effect is not only limited on the production of ATP but it effects through cellular proliferation, migration and growth factor and production of cytokines through the activation of transcription factor by ROS.

Light photons contain enormous amount of energy and its effects on cellular and Tissular level mechanism vary from clinical application of LLLT. Despite of the great amount of study and research on LLLT, exact mechanism of LLLT is still not well known, and this topic remains as the contradictory since the invention of it.
\section{Light sources in LLLT}
Choice between LED and laser light for LLLT therapy is also the one of the topic of debate. He-Ne lasers were pioneer source for LLLT, but semiconductors laser such as gallium arsenide (GaAS) lasers has also gained wide interest. As discussed above, use of non coherent light emitted from LED has also been found useful for therapeutic purposes.  The new emerging non-coherent light source is the organic light emitting diodes (OLEDs), these consists of emissive electroluminescent layer is made of organic compound which emit light in response to electric current \cite{xiao2011recent, da2006collagen}.The mechanism of working of these OLED is the same as traditional semiconductor material, electron and holes combine to make exciton. Radiations are emitted as a result of the decay of the excited state and the wavelength of the emitted radiations fall in the visible regions.

As discusses in previous section effective wavelength for LLLT falls within the optical window at which the effective penetration is maximized (Fig. \ref{fig:prism10}). Effective wavelengths for vary from 600-1070 nm, as the principle chromophores hemoglobin and melanin have high absorption power at wavelength shorter than 600 nm. Superficial tissues are treated at the wavelengths of 900-700 nm, whereas 780-950 nm wavelengths are used to treat deeper seated tissues. Wavelengths in the range of 700-770 have shown low bio activity so these wavelengths are not much used in LLLT. Some reports suggested the effective wavelength for CCO in the IR or Near IR region by carbon dioxide laser (10.6 μm) and broad band IR sources in the range of 10–50 μm \cite{xiao2011recent, hoffmann2007principles, pinheiro2005polarized, ribeiro2004effects, castano2007low}. Water in nano structured form is mainly the chromophores in these situations which exist in biological membranes. It is still not clear which wavelength is suitable or effective at which CCO absorption stops and water absorption starts \cite{santana2010interaction, santana2010theoretic}.
\section{Dosimetry and Biphasic Dose Response}
\subsection{Dosimetry}
The power of light for LLLT varies depending upon the application, but generally it lies in the range of 1-1000 mW. Effectiveness of the therapy greatly depends upon the energy and power density used and it has both upper and lower threshold limit at which LLLT is effective. Beyond this limit it has too much power to have any beneficial effect or too weak to have any effect on cells or tissues.
Some insights about the effectiveness of different paramteres (wavelenght, coherence, polarization etc) on LLLT has been given in the previous section.  Table 2 summarizes the parameter for dosimetry. Table 3 illustrate the  parameters for the dose rate \cite{chung2012nuts}.

\begin{table}[h]
\centering
\caption{Radiation parameters \cite{chung2012nuts}}
\label{my-label}
\begin{tabular}{|l|l|l|}
\hline
\textbf{Parameter}       & \parbox[c]{2.5cm}{\raggedright \textbf{Unit}}                                                                                   & \parbox[c]{6.5cm}{\raggedright \textbf{Description}}                                                                                                                                                                                                                                                                                        \\ \hline
\textbf{Wavelenght}      & \parbox[c]{2.5cm}{\raggedright nm}                                                                                              & \parbox[c]{8.5cm}{\raggedright Light has wave like property due to its,electromagnetic nature. Effective Light wavelenght for LLLT suggested to,exist at range of 600-1000 nm as suggested by clinical trials and peaks of,cytochrome c oxidase in that range.}                                                                            \\ \hline
\textbf{Irradiance}      & \parbox[c]{2.5cm}{\raggedright $W/cm^2$ }                                                                         & \parbox[c]{8.5cm}{\raggedright Rate at which energy fall on a unit,surface and is calculated as Power (W)/Area ($cm^2$) = Irradiance}                                                                                                                                                                                                         \\ \hline
\textbf{Pulse Structure} & \parbox[c]{2.5cm}{\raggedright Peak Power (W), Pulse Width(s), Pulse frequency (Hz)} & \parbox[c]{8.5cm}{\raggedright Pulse,beam is found to be more effective than continuous wave (CW), so if the beam,if pulsed than the power should be average power as is calculated as; Peak,Power (W) × pulse width (s) × pulse frequency (Hz) = Average Power (W).,Optimal frequency and pulse duration is still need to be determined.} \\ \hline
\textbf{Coherence}       & \parbox[c]{2.5cm}{\raggedright Depend,upon spectral band width}                                                                 & \parbox[c]{8.5cm}{\raggedright Speckle,which produces photobiomodulation interaction in cells and orgenelles,originate by lasers. The dimension of orgenlle such as mitochondria falls,simultaneouly with speckle dimension.}                                                                                                              \\ \hline
\textbf{Polarization}    & \parbox[c]{2.5cm}{\raggedright Linear,or circular polarization}                                                                 & \parbox[c]{8.5cm}{\raggedright Effects,of polarized differ from the linear light. But it seems to scatter in highly,scattered media such as tissue in the first few hundred μm. Some,study reported the effectiveness of polarized light in the wound healing and,burns \cite{xiao2011recent, hoffmann2007principles, huang2009phantom}.}                                                   \\ \hline
\end{tabular}
\end{table}

\begin{table}[h!]
\centering
\caption{Dose calculation (time/energy/fluence delivered) \cite{chung2012nuts}}
\label{my-label}
\begin{tabular}{|r|r|r|}
\hline
 \textbf{Parameter}\centering &  \textbf{Unit}          & \parbox[c]{6.5cm}{\raggedright \textbf{Description}}
 \\ \hline
\textbf{Energy}  & \parbox[c]{2.5cm}{\raggedright Joule (J)}              & \parbox[c]{8.5cm}{\raggedright It is calculated as Power (W) × time (s) = Energy,(Joules).it ignores the irradiance by mixing medicine and dose into single expression.Joule cannot be assumed as an expression for dose because it assumes inverse relationship between,irradiance and time.} \\ \hline
\textbf{Energy Density} & \parbox[c]{2.5cm}{\raggedright $J/cm^2$} & \parbox[c]{8.5cm}{ Energy density is the common expression for LLLT, but it is still unreliable expression for dose due to the above stated reason.}
\\ \hline

\textbf{Irradiation time}& \parbox[c]{2.5cm}{\raggedright Seconds}                & \parbox[c]{8.5cm}{\raggedright In vitro g the above problems it is reliable to record,and prescribe LLLT irradiation parameter, Irradiation time is defined a time,at which energy fall on a unit surface area.}
\\ \hline
\textbf{Treatment interval}& \parbox[c]{2.5cm}{\raggedright Hours, days or weeks}   & \parbox[c]{8.5cm}{\raggedright The effect of treatment interval depends upon the,treatment part. Study suggest aside from some acute burns treatment time take,weeks to months treatment interval to achieve any appreciable effect.}

 \\ \hline
\end{tabular}
\end{table}
The parameters in table 2 are largely inter related to each other, which showed there is not much study reported on the effect of individual parameters and it is unlikely to conduct this study due to the complexity of the method. Due to this reason parameter’s choice is dependent upon guess based upon  practitioner’s experience or personal choice instead of guidelines.
\subsection{Biphasic Dose response}
As it is established there is the optimal or threshold limit of irradiance and time limit for stimulating a response in LLLT beyond of which limit either response of radiation is inhibited or no response reported at all \cite{castano2007low, haxsen2008relevance, lanzafame2007reciprocity}.

 Hugo Schulz \cite{lanzafame2007reciprocity, mester1985biomedical, sommer2001biostimulatory} in 1887 showed simulatory effect of different poisons on the metabolism of yeast at low doses. Rudolph Arndt \cite{martius1923arndt} further proposed slight acceleration in activity at weak stimuli, further increase in activity at stronger stimuli and at peak stronger stimulus inhibit activity. This is also known as Arndt-Schulz Law which can summarissed as “For every substance, small doses stimulate, medium doses inhibit and large doses terminate \cite{fisher2003does}”. The other known term in science and medicine to relate this phenomenon is Hueppe’s Rule or hormesis.  Ferdinand Hueppe in 1896 showed low dose stimulation and high dose inhibition of bectaria by toxic agent inspiring from the Hugo Schulz idea \cite{martius1923arndt, hayworth2010vivo, huang2009biphasic}.

\begin{figure}[h!]\centering
\includegraphics[scale=0.9]{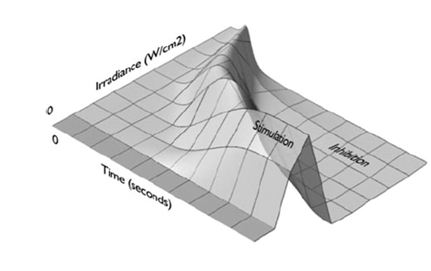}
\caption{3D plot representing the biphasic response in LLLT at different combinations of irradiance and time \cite{huang2009biphasic}}\label{fig:prism13}
\end{figure}

Fig \ref{fig:prism13} showed the variation in response of LLLT at different irradiance and time in 3D plot to show the different biphasic reponse at various irradiance (medicine) time or fluence.

\begin{figure}[h!]\centering
\includegraphics[scale=0.7]{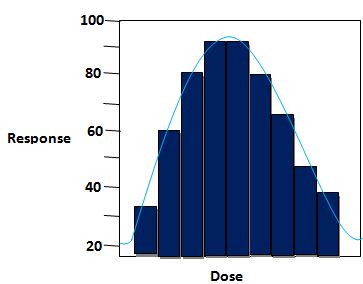}
\caption{Idealized dose response curve of biphasic response \cite{huang2009biphasic}}\label{fig:prism14}
\end{figure}

In conclusion in order to describe  the dose response to light at tissular,cellular or clinical level biphasic curve is used (idealized illustration is shown is fig \ref{fig:prism14}). It suggest how much energy is sufficient, insufficient or excessive to stimulte, inhibit or kill the dose response.
\section{Application Techniques for LLLT}
\begin{figure}[h!]\centering
\includegraphics[scale=0.7]{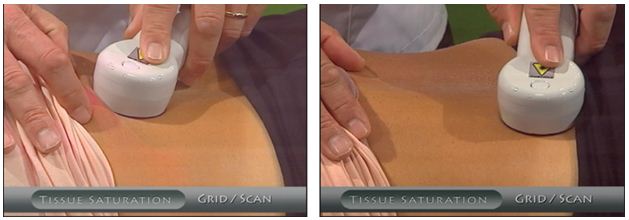}
\caption{Application of Tissue saturation technique for the treatment of paraspinal muscle. Courtesy of Multi Radiance Medical \cite{kneebone2006practical}}\label{fig:prism15}
\end{figure}

Application of Laser beam on patients can be done by different techniques. There are mainly folllowing techniques curently employed by medical doctors or physicists \cite{kneebone2006practical}.
\begin{itemize}
\item Tissue saturation technique.
\item Trigger points Technique.
\item  Laser puncture technique.
\end{itemize}
Tissue saturation technique invovled the pressing the emitter or probe on the patient skin (fig. \ref{fig:prism15}) and hold a period of time so the light can penetrate to tissue and then moving back and forth for the saturation of tissue.

In order to increase the penetration of laser light into tissue pressure on the skin surface is increased by probe or emitter, in this way capillary blood in the superficial tissue is displaced and blood flow toward the treatment area is decreased. Since the photons penetration is inversely proportional to the amount of water content in the tissue, so by decreasing the blood flow toward the tretment area we can achieve photon penetration during treatment \cite{tuner2004laser}.

\begin{figure}[h!]\centering
\includegraphics[scale=0.7]{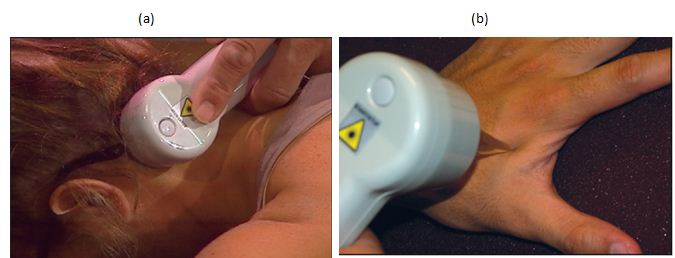}
\caption{(a) occipital TP treatment (b) Acupoint stimulation with special acupoint probe - Courtesy of Multi Radiance Medical \cite{kneebone2006practical}}\label{fig:prism16}
\end{figure}

The second technique is accomplished by stationary contact on the trigger point similar to the above treatment method (fig.\ref{fig:prism16}).

There are different types of trigger points: active or passive (latent) and primary or secondary TPs \cite{simunovic1996low}. Only active TPs ($0.5 cm^2$ in diameter) are associated with pain while rest are associated with dysfunction. Pain induced by hypersensitive TPs compression is named as referred pain.

The location of the TP can be found in all types of muscles, but mostly it is found at postural and mastication muscles. TPs are caused by the scar tissue, surgical incision, kleoids, local infection and infected tooth sockets.

Third technique is accomplished by placing emitter or probe on acupoints or special acupoint probe may be used (fig. \ref{fig:prism16}b) \cite{whittaker2004laser}. A. SCHLAGER and his coworkers \cite{schlager1998laser} applied acupuncture techniue for the treatment of onpostoperatice vomiting in children undergoing strabismus surgery. This technique significantly lower the vomiting rate than other placebo group placed for testing.

\section{Conclusion}
Use of LLLT between 1960 to 1970 found to be very useful for the wound heaing and pain type, but to the the little information about the physics and mechanism this remained to be controversial topic to use it for clinical purposes. With few years later with more information at our disposal researchers and scientist tried to answer more basic questions about the physics and beam interaction mechanism at tissular and cellular level. Many different light sources such as conventional light or LED has also been used for LLLT. In discussing the published work on the LLLT, it is shown that coherence and monochromaticty of light does not show any bonus effect as compare to conventional light sources or LED, but the use of laser is thought to be more favourable in the light of laser beam interaction with tissues and cells. Due to the scarcity of information about the effectiveness of different parameters such as wavelenght, frequency, coherence etc selection of these parameters is dependent upon pracioner’s experience. That’s why there is need of published reports on the selection of these parameters to make this the progress of this technique scientifically consistent and accepted.
\section*{Acknowledgement}
Author is immensely thankful to Dr. Malik Sajjad Mahmood for their useful comments and suggestions during manuscript preparation and Mr. AmanUllah Malik for latex help.
\bibliographystyle{ieeetr}
\bibliography{T7.Bibliography}
\end{document}